\documentclass[prl,twocolumn,showpacs,preprintnumbers,amsmath,amssymb]{revtex4}

\usepackage{amsmath,amssymb}
\usepackage{graphicx}
\usepackage{dcolumn}
\usepackage{bm}
\usepackage{color}
\usepackage{epstopdf}
\newcommand{\mr}{\mathrm}

\begin{document}
\title{Atomic clocks with suppressed blackbody radiation shift}

\author{V. I. Yudin\footnote{e-mail address: viyudin@mail.ru}, A. V. Taichenachev, M. V. Okhapkin\footnote{also PTB}, and S. N. Bagayev}
\affiliation{Institute of Laser Physics SB RAS, Novosibirsk 630090, Russia}
\affiliation{Novosibirsk State University, Novosibirsk 630090, Russia}
\affiliation{Novosibirsk State Technical University, Novosibirsk 630092, Russia}
\author{Chr. Tamm, E. Peik, N. Huntemann, T. E. Mehlst\"{a}ubler, and F. Riehle}
\affiliation{Physikalisch-Technische Bundesanstalt (PTB),
Bundesallee 100, 38116 Braunschweig, Germany}
\date{\today}

\begin{abstract}
We develop a nonstandard concept of atomic clocks where the
blackbody radiation shift (BBRS) and its temperature fluctuations
can be dramatically suppressed (by one to three orders of
magnitude) independent of the environmental temperature. The
suppression is based on the fact that in a system with two
accessible clock transitions (with frequencies $\nu^{}_1$ and
$\nu^{}_2$) which are exposed to the same thermal environment,
there exists a ``synthetic'' frequency
$\nu^{}_\mr{syn}$~$\propto$~($\nu^{}_1-\varepsilon^{}_{12}\nu^{}_2$)
largely immune to the BBRS. As an example, it is shown that in the
case of $^{171}$Yb$^+$ it is possible to create a clock in which
the BBRS can be suppressed to the fractional level of 10$^{-18}$
in a broad interval near room temperature (300$\pm 15$~K). We also
propose a realization of our method with the use of an optical
frequency comb generator stabilized to both frequencies $\nu^{}_1$
and $\nu^{}_2$. Here the frequency $\nu^{}_\mr{syn}$ is generated
as one of the components of the comb spectrum and can be used as
an atomic standard.
\end{abstract}


\maketitle

The main progress in modern fundamental metrology is connected
with the development of atomic clocks. The most promising
frequency standards today are based on single trapped ions
\cite{Rosenband08} and on ensembles of neutral atoms confined to
an optical lattice at the magic wavelength \cite{aka08,ludlow08}.
It is believed that these clocks can provide frequency references
with unprecedented small systematic uncertainties in the
10$^{-17}$-10$^{-18}$ range. This progress will probably lead to a
redefinition of the unit of Time and to new fundamental tests of
physical theories in particular in the fields of General
Relativity, cosmology, and unification of the fundamental
interactions \cite{Turyshev09, Peik08}.

The largest  effect that contributes  to the systematic
uncertainty of many atomic clocks is the interaction
of the thermal blackbody radiation with the atomic eigenstates.
This effect was first considered in 1982 for cesium atomic
clocks~\cite{Itano}, but remains up to now a major problem for
many modern atomic time and frequency standards. At present there
exist three approaches to tackle the blackbody radiation shift
(BBRS) problem. The first one is the use of cryogenic techniques
to suppress this shift to a negligible level. This approach is
pursued for the mercury ion clock \cite{Oskay}, for the Cs
fountain clock \cite{Levi2010}, and for the Sr optical lattice
clock \cite{Middelmann}. The second approach is the precise
temperature stabilization of the experimental setup in combination
with theoretical and/or semiempirical numerical calculations of
the shift at given temperature \cite{Angstmann06, Safronova10}.
The third approach is based on the choice of an atom or ion where
both levels of the reference transition have approximately the
same BBRS. Here the most promising candidate is $^{27}$Al$^+$ with
a fractional BBRS of the reference transition frequency of
$\sim$10$^{-17}$ \cite{Rosenband08,Chou}, followed by
$^{115}$In$^+$ \cite{Becker,Safronova}. However, the latter
approach limits the choice of candidates for tests of fundamental
theories.

In the present paper we propose an alternative method allowing us
to suppress the BBRS and its fluctuations in atomic frequency
standards by one to three orders of magnitude without using
cryogenic techniques and precise temperature stabilization. Our
approach is based on the use of two reference transitions in an
identical thermal environment. We show that in such a system there exists a combined frequency for which
the BBRS is significantly suppressed over a wide temperature
range. For instance, a trapped
$^{171}$Yb$^+$ ion meets this condition in a straightforward way,
because $^{171}$Yb$^+$ has at least three suitable reference
transitions: an electric-quadrupole and an electric-octupole
optical transition \cite{Tamm09,Hosaka09,Sherstov10}, and a
magnetic-dipole radiofrequency (rf) transition between the
ground-state hyperfine sublevels. Apart from laboratory standards, the proposed method can be
particularly useful in cases where it is impossible to control the
environmental temperature with sufficient accuracy or to use
cryogenic techniques, for instance in transportable frequency
standards or in space-based clocks that approach the Sun in order
to test the local position invariance underlying General
Relativity \cite{Turyshev09}.

Our approach is based on the fact that for the large majority of
transitions in atoms or ions that are of interest as frequency
standard reference transitions, the temperature dependence
$\Delta(T)$ of the BBRS is very well approximated by the law
$\propto T^4$. Consider now two clock transitions with frequencies
$\nu^{(0)}_1$ and $\nu^{(0)}_2$ exposed to the same thermal
environment, i.e., located in the same probe volume. We assume
that $\nu^{(0)}_1 < \nu^{(0)}_2$. The effect of the BBRS on each
transition frequency can be represented as:
\begin{equation}\label{omegaT}
\nu^{}_{j}(T) \approx \nu_{j}^{(0)}+a^{}_j
\left(\frac{T}{T_0}\right)^4\quad (j=1,2)\,,
\end{equation}
where $a^{}_j$ is an individual characteristic of the transition $j$
determined by the atomic structure and $T_0$ is the mean temperature of
the clock operation. Let us introduce the coefficient
$\varepsilon^{}_{12}=a^{}_1/a^{}_2$. As is easily seen, the following superposition
does not experience the BBRS: $\nu^{}_1(T)-\varepsilon^{}_{12}\nu^{}_2(T)=\nu_1^{(0)}-\varepsilon^{}_{12}\nu_2^{(0)}$.
In compliance with this we define a new ``synthetic'' frequency $\nu^{}_\mr{syn}$ as
\begin{equation}\label{omegaG}
\nu^{}_\mr{syn}=R[\nu^{}_1(T)-\varepsilon^{}_{12}\nu^{}_2(T)]
=R[\nu_1^{(0)}-\varepsilon^{}_{12}\nu_2^{(0)}]\,,
\end{equation}
where $R$ is some numerical multiplier whose value can be chosen
freely. Thus, one can use the frequency $\nu^{}_\mr{syn}$ as a new
clock output frequency which is immune to the BBRS and to
fluctuations in the operating temperature, while the thermal
shifts $a^{}_jT^4$ of the working frequencies $\nu_j$ can be
large.

\begin{figure}[t]
\centerline{\scalebox{0.4}{\includegraphics{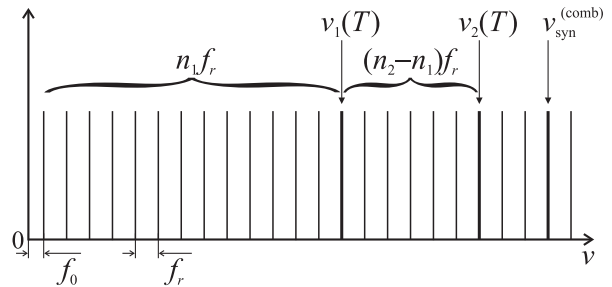}}}\caption{Illustration of femtosecond comb stabilized to the two clock
transitions with frequencies $\nu^{}_1$ and $\nu^{}_2$ at
a given temperature $T$.}\label{comb_fig}
\end{figure}

One possibility is to independently measure both frequencies
$\nu^{}_{1,2}(T)$ and to use for the clock operation the synthetic
frequency according to Eq. (\ref{omegaG}) (assuming, for example,
$R=\pm 1$).  In this case, obviously the synthetic frequency does
not directly correspond to any frequency of a physical signal.
Another approach is to synthesize this frequency as a real
physical signal by means of an optical frequency comb generator.
Let us consider the situation where two modes of the frequency
comb generator are stabilized to the two optical frequencies
$\nu^{}_{1,2}(T)=f^{}_0+n^{}_{1,2}f_{r}$ at a given temperature
$T$ (see Fig.\ref{comb_fig}). As a result, the parameters of the
comb spectrum, i.e., the pulse repetition rate $f_{r}$ and the
offset frequency $f_0$ are unambiguously determined and the
frequency of the $m$-th mode equals:
\begin{eqnarray}\label{m_mode}
\nu_m(T)&=&f_0+mf_{r}=\nonumber\\
&&\frac{m(\nu^{(0)}_2-\nu^{(0)}_1)+n^{}_2\nu^{(0)}_1-n^{}_1\nu^{(0)}_2}{n^{}_2-n^{}_1}+
\\
&&\frac{m(a^{}_2-a^{}_1)+n^{}_2a^{}_1-n^{}_1a^{}_2}{n^{}_2-n^{}_1}\left(\frac{T}{T_0}\right)^4\,.\nonumber
\end{eqnarray}
From this expression one can define a number $m=m^{}_0$ for
which the coefficient of the temperature-dependent term is zero:
\begin{equation}\label{m0}
m^{}_0=\frac{n^{}_1a^{}_2-n^{}_2a^{}_1}{a^{}_2-a^{}_1}
=\frac{n^{}_1-\varepsilon^{}_{12}n^{}_2}{1-\varepsilon^{}_{12}}\,.
\end{equation}
This shows that the BBRS is
suppressed for the frequency $\nu_{m_0}$. After a simple transformation we see that
the frequency $\nu_{m_0}$ is the synthetic frequency
defined in Eq. (\ref{omegaG}),
\begin{equation}\label{omega_m0}
\nu^{}_{m_0}=\nu^\mr{(comb)}_\mr{syn}=\frac{\nu_1^{(0)}-\varepsilon^{}_{12}\nu_2^{(0)}}{1-\varepsilon^{}_{12}}\,,
\end{equation}
as it should be.
Here, $m_0$ is the natural number closest to the value of the right-hand side of Eq. (\ref{m0}).
For this it is necessary to satisfy the condition $m^{}_0>0$ that is
equivalent to $\nu^\mr{(comb)}_\mr{syn}>0$ in Eq. (\ref{omega_m0}).

Apart from the frequency $\nu_{m_0}$ which is a component of the
optical spectrum of the frequency comb generator, in our system
one can also define the much smaller frequency
\begin{equation}\label{omega_low}
\frac{\nu_{m_0}}{m^{}_0}=f_{r}+\frac{f^{}_0}{m^{}_0}\,
\end{equation}
which corresponds to a rf standard at $\nu_{m_0}$/$m^{}_0$. Since
the frequencies $f_{r}$ and $f^{}_0$ can be extracted from a
stabilized comb generator with negligible error, one can use them
to synthesize $\nu_{m_0}$/$m^{}_0$. This synthesized
radiofrequency has the same immunity to BBRS as $\nu_{m_0}$. It is
interesting to note that the radiofrequency given in Eq.
(\ref{omega_low}) is well-defined in our system even if $m^{}_0<0$
in Eq. (\ref{m0}), i.e., if the basic frequency component
$\nu_{m_0}$ exists only virtually.

As was shown above, in the case of a frequency comb stabilized to
two BBR-shifted clock transitions with frequencies $\nu^{}_1(T)$
and  $\nu^{}_2(T)$, there exists a frequency component $\nu_{m_0}$
(for $m^{}_0>0$) for which the thermal shift and the sensitivity
to temperature fluctuations vanish. This frequency component can
serve as an atomic frequency standard. In practice, the BBRS is
strongly suppressed for a range of comb frequencies around
$\nu_{m_0}$. The residual shift of frequency components
$\nu_{m_0\pm l}$ near $\nu_{m_0}$ is given by:
\begin{eqnarray}\label{omega_pm}\nonumber
\nu^{}_{m_0\pm l}&=&\nu_{m_0}\pm lf_{r} \\
                    &=&\nu_{m_0}\pm l\frac{\nu^{(0)}_2-\nu^{(0)}_1}{n^{}_2-n^{}_1}\pm
l\frac{a^{}_2-a^{}_1}{n^{}_2-n^{}_1}\left(\frac{T}{T_0}\right)^4.
\end{eqnarray}
This indicates that the suppression is effective as long as
$(n^{}_2-n^{}_1)\gg l$. For example, for frequencies $\nu^{(0)}_1$
and $\nu^{(0)}_2$ in the optical range, the comb mode index
difference $(n^{}_2-n^{}_1)$ will typically be in the range of
10$^5$ or higher (see the discussion for the case of
$^{171}$Yb$^+$ below). On the whole, the choice of the exact value
of the synthetic frequency $\nu^\mr{}_\mr{syn}$ is to some extent
arbitrary if one takes into account that the coefficient
$\varepsilon^{}_{12}$ is only known with limited accuracy and that
Eq. (1) is an approximation that neglects higher-order terms in
the temperature dependence of the BBRS \cite{Farleywing}.
Including higher-order terms the BBRS can be expresssed as:
\begin{equation}\label{DT}
\Delta^{(j)}(T)=a^{}_j\left(\frac{T}{T_0}\right)^4+b^{}_j
\left(\frac{T}{T_0}\right)^6+...\quad (j=1,2)\,.
\end{equation}
From this we can estimate a basic limitation of the possibility to
suppress the BBRS and its temperature dependence. Usually, near
room temperature $T_0=300$~K the contribution of the higher terms
[$b^{}_j (T/T_0)^6+...$] is a factor of 10$^{}$ to 10$^{3}$
smaller than that of the main $T^4$-term
\cite{Itano,Pal'chikov}. This indicates that here it would not be
useful to suppress the $T^4$-dependence of $\nu^{}_\mr{syn}$ to better
than one to three orders of magnitude because higher-order
contributions to the BBRS remain uncompensated. For example, in
order to achieve a suppression factor of 10$^2$ for the
$T^4$-dependence, it would be sufficient to know the coefficient
$\varepsilon^{}_{12}$ with relative uncertainty of 10$^{-2}$.

It may be noted that apart from theoretical calculations the
coefficient $\varepsilon^{}_{12}$ can be determined by purely
experimental means. To do this we can apply a quasistatic electric
field (or the field of an infrared laser) to determine the shifts
of the reference transition frequencies $\nu^{}_1$ and $\nu^{}_2$
due to the differences in the static polarizabilities of the
involved atomic energy levels. From a practical point of view it
is very advantageous that we do not need to know the magnitude of
the electric field at the place of the atoms because we only have
to determine the ratio $a_1/a_2$. If a frequency comb generator is
stabilized to  $\nu^{}_1$ and $\nu^{}_2$ as shown in Fig.1, the
frequency $\nu_{m_0}$ can be identified in a direct way as the
frequency component which does not experience any scalar Stark
shift in the applied quasistatic field.

\begin{figure}[t]
\centerline{\scalebox{0.4}{\includegraphics{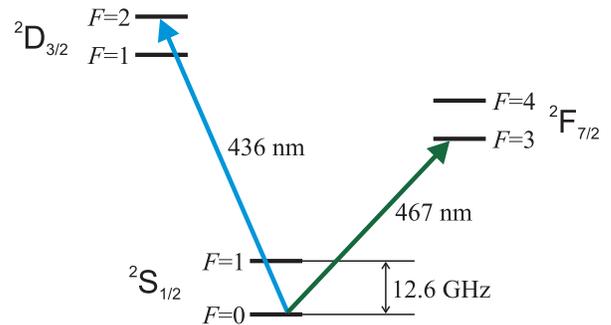}}}\caption{(Color
online) Section of the energy level scheme of $^{171}$Yb$^+$,
showing the hyperfine levels of the $^2$S$_{1/2}$ ground state and
the two lowest-lying excited states, which are metastable.
Hyperfine splittings are not drawn to scale.}\label{Yb_fig}
\end{figure}

As an example that permits the practical realization of the ideas
presented above, we consider the ion $^{171}$Yb$^+$. As shown in
Fig.\ref{Yb_fig}, the level system of $^{171}$Yb$^+$ provides two
narrow-linewidth transitions from the ground state in the visible
spectral range which can be used as reference transitions of an
optical frequency standard: the quadrupole transition
$^2$S$_{1/2}(F=0) \to ^2$D$_{3/2} (F=2), \lambda \approx 436$~nm
and the octupole transition $^2$S$_{1/2}(F=0) \to
^2$F$_{7/2}(F=3), \lambda\approx 467$~nm. More detailed
information on the spectroscopy of these transitions can be found
in \cite{Schneider,Tamm09,Hosaka09,Sherstov10}. It may be noted
that the case of $^{171}$Yb$^+$ is especially attractive because
here both clock transitions lie in a technically convenient
frequency range and experience exactly the same thermal
environment if probed in one ion.

The BBRS of the quadrupole and octupole transitions of Yb$^+$ were
calculated in Ref. \cite{Lea06}. This calculation is based on
calculated oscillator strengths and experimental lifetime and
polarizability data. The room-temperature BBRS of the quadrupole
transition is calculated as $a_\mr{quad} = -0.35(7)$~Hz
(fractional shift $5.1(1.1) \times 10^{-16}$) and that of the
octupole transition as $a_\mr{oct} = -0.15(7))$~Hz (fractional
shift $2.4(1.1) \times 10^{-16}$). The relatively small value of
$a_\mr{oct}$ and the large relative uncertainty is due to nearly
equal shifts of the $^2$S$_{1/2}$ and $^2$F$_{7/2}$ levels.

Using the results of Ref. \cite{Lea06}, for $^{171}$Yb$^+$ we find
that $\varepsilon^{}_{12} = a_\mr{oct}$/$a_\mr{quad} = 0.43(22)$.
We expect that the large uncertainty of this value can be reduced
to less than 1\% by improved atomic structure calculations or by a
direct measurement of $\varepsilon^{}_{12}$ as discussed above. In
the following, we will not take into account the present
uncertainty because our conclusions remain qualitatively unchanged
for all values of $\varepsilon^{}_{12}$ in this uncertainty range.
In particular, we find the synthetic frequency
$\nu^\mr{(comb)}_\mr{syn}\approx 607$~THz, corresponding to a
wavelength $\lambda_\mr{syn}\approx 494$~nm. This frequency lies
sufficiently close to the initial reference transitions at 436~nm
and 467~nm that it can be generated as a spectral component of a
femtosecond comb generator that is locked to the reference
transitions as shown in Fig.~1.

The higher-order contributions in Eq.(\ref{DT}) to the BBRS of the
octupole reference transition are negligible compared to that of
the quadrupole transition. For the latter, we find $b/a\approx
0.1$ at $T_0=300$~K. As a result, we estimate that the BBRS can be
suppressed to the fractional level of $2.7\times 10^{-17}$ at 300
K with variations at the level of $\pm 5 \times 10^{-18}$ if the
ambient temperature varies in a broad interval of $\pm 15$~K.

It is also possible to estimate the frequency interval around
$\nu^\mr{(comb)}_\mr{syn}$ where the components of the comb
spectrum have a similar level of suppression of the thermal shift
and of its fluctuations. For a suppression factor of 10$^2$, this
interval has a width of the order of 1000~GHz. Thus, for $d\sim
100$~MHz, the indicated interval contains 10$^4$ comb modes, each
of which could be used as a stable frequency reference.

Other variants of BBRS-free optical frequency standards at a
synthethic frequency can be conceived based on transitions
$^1$S$_0 \to ^3$P$_0$ in alkaline-earth-like neutral atoms
confined in an optical lattice. Consider, for instance, the
combination of the reference transitions of strontium ($\nu^{}_1
\approx 429$~THz, $\lambda \approx 698$~nm) and ytterbium
($\nu^{}_2 \approx 518$~THz, $\lambda \approx 578$~nm). Using the
calculations in Ref. \cite{Porsev}, in this case we obtain
$\varepsilon^{}_{12}\approx 1.69$ and an estimated synthetic
frequency $\nu^\mr{(comb)}_\mr{syn}\approx 648$~THz
($\lambda_\mr{syn} \approx 463$~nm). The technical realization of
this variant requires the operation of two lattice-based clocks
with different atoms (Sr and Yb) in the same vacuum chamber.

So far we have considered examples where both reference transition
frequencies lie in the optical region. However, it is also
possible to realize schemes where the high frequency $\nu^{}_{2}$
is optical, but the lower frequency $\nu^{}_{1}$ corresponds to a
fine- or hyperfine-structure splitting so that it lies in the
terahertz or microwave range. In contrast to the above example of
ion Yb$^+$, which seems unique because it provides two optical
reference transitions, in this case one can find many appropriate
schemes that use a single atomic species. Also the ion
$^{171}$Yb$^+$ offers the possibility of using the ground-state
hyperfine transition $F=0\to F=1$ at $\nu^{}_1$=12.6~GHz (see
Fig.\ref{Yb_fig}) as a low-frequency reference transition. For the
combination with the octupole transition $^2$S$_{1/2}(F=0) \to
^2$F$_{7/2}(F=3)$ at $\nu^{}_2$=642~THz, a numerical estimate
yields $\varepsilon^{}_{12}$$\approx$6.6$\times$10$^{-5}$ and a
synthetic frequency
$\nu^\mr{}_\mr{syn}=-(\nu^{}_1-\varepsilon^{}_{12}\nu^{}_2)\approx$30~GHz.
(We expect that our estimate on $\varepsilon^{}_{12}$ is accurate
to $\pm 20$~\%, but it is not principal for further results.) Here
the $T^6$-contribution to the BBRS (see Eq.(\ref{DT})) limits the
BBRS suppression at the fractional level of 7.5$\times$10$^{-19}$ at
$T$=300~K with variations of $\pm$2$\times$10$^{-19}$ in the
temperature interval of 300$\pm 15$~K. However, we should also
take into account the shift of the ground-state
hyperfine levels ($\propto T^2$) due to the magnetic blackbody radiation field \cite{Itano}.
The corresponding BBRS of the hyperfine frequency $F=0\to F=1$ for $^{171}$Yb$^+$ is:
\begin{equation}\label{MD}
\Delta^{(1)}_\mr{magn}(T)=-1.616\times
10^{-7}\times\left(\frac{T(\mr{K})}{300}\right)^2\; \mr{Hz}.
\end{equation}
For $\nu^\mr{}_\mr{syn}$=30~GHz this shift
results in a fractional level of
5.4$\times$10$^{-18}$ at $T$=300~K with a variation of
$\pm$5$\times$10$^{-19}$ for 300$\pm 15$~K. Since the magnetic
BBRS contribution can be readily calculated with an accuracy of
less than 1\%, it is possible to apply a corresponding
correction to $\nu^\mr{}_\mr{syn}$ with an uncertainty
contribution of less than 10$^{-19}$.

For $^{171}$Yb$^+$ we have thus shown the possibility to create a
synthetic-frequency-based atomic clock with a fractional
uncertainty contribution due to BBRS of $<$1.5$\times$10$^{-18}$
in a broad interval of 300$\pm 15$~K. To achieve this, we need to
know the coefficient $\varepsilon^{}_{12}$ with a relative
accuracy in the range of 0.1-0.2\%. In order to
reduce the BBRS uncertainty contribution to less than 10$^{-17}$,
the value of $\varepsilon^{}_{12}$ needs only be known with a
relative accuracy of 3\%. We also have pointed out that for
$^{171}$Yb$^+$ the combination of the octupole optical clock
transition with the ground-state hyperfine transition can yield a
much better BBRS suppression than the combination of the octupole
and quadrupole optical clock transitions. The use of the
quadrupole transition yields a lower BBRS suppression because the
upper level $^2$D$_{3/2}$ is connected to the $^2$P$_{1/2}$ level
by a strong infrared transition at 2.44~$\mu$m, which produces a
relatively large $T^6$-contribution to the BBRS. The final
comparison of the two options for BBRS suppression should also
include detailed estimates on the magnitudes of other systematic
uncertainty contributions in the considered experimental setup.

The concept of a synthetic atomic frequency standard based on two
reference transitions can also be extended to the case that both
reference frequencies $\nu{}_{1,2}$ lie in the microwave range.
Atomic fountain clocks are based on reference transitions in the
microwave range between the ground-state hyperfine sublevels of
alkali atoms. For a synthetic atomic fountain frequency standard,
for instance the combination $^{87}$Rb ($\nu^{}_1 \approx
6.8$~GHz) and $^{133}$Cs ($\nu^{}_2 \approx 9.2$~GHz) can be
considered. Here, at the synthetic frequency
$(\nu^{}_1-\varepsilon^{}_{12}\nu^{}_2)\approx 1.9$~GHz it is
possible to suppress the fractional BBRS of the individual
standards by two orders of magnitude. It is interesting to note
that nearly optimal conditions for the efficient suppression of
the BBRS are realized in the dual Rb$/$Cs fountain clock described
in Ref. \cite{Guena} because here both reference transitions are
exposed to the same thermal environment.

We finally note that the $^{171}$Yb$^+$ optical frequency standard
is a very sensitive system for a search for temporal variations of
the fine structure constant $\alpha$ \cite{Dzuba,Lea}. The
frequencies of the quadrupole and octupole reference transitions
of Yb$^+$ have significant contributions from relativistic effects
and would undergo changes with different sign in consequence of a
change of $\alpha$. The synthetic frequency that eliminates the
BBRS retains this sensitivity. The $\alpha$-dependence of an
atomic transition frequency may be expressed generally as
$\nu=\nu_0 + qx$, where $x\equiv (\alpha/\alpha_0)^2-1$, $\nu_0$
defines the frequency at the present value of the fine structure
constant, $\alpha_0$, and $q$ quantifies the sensitivity to
changes of $\alpha$ \cite{Dzuba}. The $q$ parameter for the
synthetic frequency is simply given by
$q_\mr{syn}=(q_1-\varepsilon^{}_{12}q_2)/(1-\varepsilon^{}_{12})$.
For Yb$^+$, with $q$ parameters as given in Ref. \cite{Dzuba},
$q_\mr{syn}$ amounts to about -3220 THz. Comparison with the Yb$^+$
synthetic frequency $\nu^\mr{(comb)}_\mr{syn}\approx 607$~THz
indicates the strong sensitivity. In a test for variations of
$\alpha$, the synthetic frequency would have to be compared to an
``anchor'' reference transition with small $q$ value, like the
$^1$S$_0\rightarrow ^3$P$_0$ transition in Al$^+$.

In conclusion, we have proposed and developed the concept of an
atomic frequency standard where the frequency shift due to the
ambient blackbody radiation and related fluctuations of the output
frequency can be suppressed by one to three orders of magnitude
without using cryogenic techniques. We also expect that our
results will stimulate refined atomic structure calculations on
Yb$^+$ and other atomic systems that are of interest in this
context. Such calculations can yield precise values for the
frequency synthesis parameter $\varepsilon^{}_{12}$ and determine
limitations of the achievable BBRS suppression.

We thank U. Sterr for useful discussions. This work was supported
by QUEST, DFG/RFBR (grant 10-02-91335), RFBR (grant Nos.
10-02-00406, 11-02-00775, and 11-02-01240),  RAS, Presidium SB
RAS, and by the federal programs ``Development of scientific
potential of higher school 2009-2010'' and ``Scientific and
pedagogic personnel of innovative Russia 2009-2013''.

\end{document}